# Hierarchical Linear Modeling Approach to Measuring the Effects of Class Size and Other Classroom Characteristics on Student Learning in an Active-Learning Based Introductory Physics Course

Connor Gorman*, David Webb, Kevin Gee

University of California, Davis

*cogorman@ucdavis.edu

The effect of class size on student learning has numerous policy implications and has been a major subject of conversation and research for decades. Despite this, few studies have been done on class size in the context of university settings or physics courses. After discussing some of the reasoning behind hierarchical linear modeling (HLM) as well as how to interpret the results of an HLM analysis while grounding this study in measurement theory as it applies to course grades, this paper goes on to examine the effect of class size in active-learning based introductory physics courses using a series of hierarchical linear models. It is found that class size over the ranges studied does not have a significant effect on student grades which were used as a proxy for students' understanding of the underlying material. However, a variety of issues and limitations means this is certainly not the end of the story and there is still much to be done and discussed when it comes to these courses and the various factors which affect student achievement in them, both theoretically and empirically.

# Motivation and Background

The effect of class size on student learning (sometimes referred to as achievement or understanding) has been a hotly contested and heavily researched topic for quite some time now. The general consensus is that when implemented properly classes of at most 18 students are optimal, at least in the early grades (often Kindergarten through third grade) [1]. However, there is also evidence (using multilevel modeling) to suggest that the effect of class size may actually vary depending on other factors [2]. This implies that more research is necessary to determine the conditions under which class size is relevant and the different ways that its effects manifest themselves depending on the circumstances. It would therefore be beneficial to conduct such research in educational settings that have not yet been studied in the context of class size. One of these areas is secondary and post-secondary education and another is physics courses. Both of these have seen relatively few studies on the issue of class size, though there are a few exceptions yielding mixed results [3, 4, and 5]. This means that such studies could be of use to the education and Physics Education Research (PER) communities both to progress the knowledge of student learning in these settings and to help inform where physics educators should focus their limited attention and resources.

Furthermore, physics education is moving in the direction of taking a more active approach to learning [6] and even with technological advancements it is likely that for active-learning techniques to work teachers will still have to be physically present [7]. It also seems reasonable to speculate that class size would have an even greater impact in an active-learning environment than a more traditional one since when students passively watch someone lecture the number of students present is less likely to impact a given student's understanding of the material than it would if students are working together on problems with guidance from the instructor. Essentially, the number of students present has a higher chance of affecting classroom dynamics in the latter case than the former. Therefore, it is especially desirable to study the effect of class size in the context of physics courses that employ active-learning techniques. Statistically, this speculation would manifest in measures of active-learning moderating the effect of class size on student understanding, though at least one study has found no significant interaction between a school's average class size and certain measures of active-learning when it comes to overall science achievement [8]. However, overall science achievement as measured by a single standardized test in secondary school is not necessarily the same as student understanding of physics content in an introductory college course.

Given all of the above, the main research question for this study is: "What is the relationship between class size and students' understanding of physics concepts in active-learning based introductory physics courses?" While the primary focus here is on class size, other class (and individual) level variables like student GPA (individual level) and class start times (class level) will also be included in the analysis as controls which will provide an opportunity to examine their effect on students' understanding of these concepts as well.

# Hierarchical Linear Modeling

The main statistical method used in this study is hierarchical linear modeling (HLM), sometimes referred to as multilevel modeling. This is because there are many situations where data points can be naturally grouped together (clustered) as a result of the larger structures that encompass them. One of the most straightforward examples of this occurs in formal schooling where students are grouped into (or in other words, clustered into and nested within) classes which are themselves nested within schools or universities. HLM considers each of these sequential groupings to be a new level, where lower levels are nested within higher levels, and accounts for this nesting structure by associating each variable (and corresponding data) with the appropriate level. In the above example, level 1 would be the student level so all of the variables and data describing students' individual traits (such as their genders and test scores) would be associated with level 1 while classes constitute level 2, meaning any variables and data that have to do with the class that a given student is in (such as the teacher's experience and how many students are in the class) would exist at level 2, and schools or universities (along with their associated variables and data, such as median parental income and location) make up level 3.

On the other hand, standard regression techniques do not account for this nesting structure which can cause a variety of problems. First off, standard regression techniques lead to a higher chance of identifying false positives, effects that seem significant in the sample and analysis when they really are not significant in the overall population, than would be the case using HLM. This is because individual observations and data from levels above level 1 are not all independent and yet, standard regression techniques naturally treat them as if they are [9, p. 3]. For instance, the amount of experience that a teacher has will be the same for all of the students in their class (in a given year) and is not an independent observation for each of these students despite standard regression techniques typically treating these data points as if they were independent observations. It is known that incorrectly treating these observations as independent results in narrower confidence intervals that could cause an otherwise non-significant effect to appear significant [9, p. 3]. Conceptually, there is also a higher risk of incorrectly drawing conclusions about one level in a nested hierarchy based on analyses involving a different level using standard regression techniques [9, p. 3-4].

Finally, HLM allows useful information to be garnered that would not be accessible using standard regression techniques. In particular, it allows the intercepts for a given outcome (dependent) variable to be different for different clusters and regression coefficients (slopes) on independent variables at lower levels to be different for different higher level clusters. For example, if student grades are being used as the outcome variable in a three level situation involving students (level 1), classes (level 2), and schools (level 3) then there is only one intercept when a standard regression analysis is performed but using HLM it is possible to find an overall intercept as well as an intercept for each class and an intercept for each school. Similarly, when a standard regression analysis is performed there is a single slope associated with each predictor (independent) variable whereas using HLM it is possible to associate separate slopes for each class and/or school with each level 1 predictor variable (so things like gender and test scores) and it is also possible to associate separate slopes for each school with each level 2 predictor variable (so things like teacher experience and class size).

While there are more advanced HLM techniques, the most basic method requires the outcome variable to be continuous (at least interval level as opposed to categorical or ordinal) and at level 1 (which is often, though not always, the level of individuals). If this is the case then the general three level HLM model describing such a situation can be written as:

Level 1:

$$\mathrm{Outcome}_{ijk} = \beta_{0jk} + \beta_{1jk} * \mathrm{Level1Variable1}_{ijk} + \ . \ . \ . \ + \beta_{Mjk} * \mathrm{Level1VariableM}_{ijk} + \varepsilon_{ijk}$$

Level 2:

$$\beta_{mjk} = \gamma_{m0k} + \gamma_{m1k} * \mathrm{Level2Variable1}_{jk} + \ . \ . \ . \ + \gamma_{mNk} * \mathrm{Level2VariableN}_{jk} + u_{mjk}$$

$$\forall m \in \{0, \ . \ . \ ., \ M\}$$

Level 3:

$$\gamma_{mnk} = \pi_{mn0} + \pi_{mn1} * \mathrm{Level3Variable1}_{k} + \ . \ . \ . \ + \pi_{mnP} * \mathrm{Level3VariableP}_{k} + \nu_{mnk}$$

$$\forall m \in \{0, \ . \ . \ ., \ M\}, \forall n \in \{0, \ . \ . \ ., \ N\}$$

Here i is an index that labels the entities where level 1 data is coming from (such as students), j is an index that labels the entities where level 2 data is coming from (such as classes), and k is an index that labels the entities where level 3 data is coming from (such as schools). There are M level 1 variables, N level 2 variables, and P level 3 variables.

Notice how the basic mathematical premise here is that intercepts and regression coefficients are not only allowed to vary in the ways discussed previously but that they can also be modeled and interpreted as functions of (higher level) predictor variables. The above equations give a description of the general three level situation in terms of levels. It is also possible to write a composite description of HLM equations by taking the (higher level) coefficient equations and plugging them into their corresponding terms in lower level equations but doing so can get complicated while not being all that informative when the model that is under consideration has few, if any, interaction terms so this paper will be sticking to the levels description.

One of the first aspects of the above equations which may seem unfamiliar are the $\varepsilon_{ijk}, u_{mjk},$ and $\nu_{mnk}$ terms. These are error terms. Essentially, they are the difference between actual values of the outcome variable and their corresponding predicted (from the predictor variables) values at different levels (when m and n are 0) or the difference between the regression coefficient associated with a given level 1 predictor variable, a given level 2 predictor variable, or an interaction term between a given level 1 predictor variable and a given level 2 predictor variable for different level 2 or level 3 clusters (when m and/or n are not 0). $\varepsilon_{ijk}$ might seem somewhat familiar since it is similar to the error term in standard regression except instead of being the difference between the actual value of the outcome variable for a given observation and the overall predicted value of the outcome variable for that observation it is the difference between the actual value of the outcome variable for a given observation (labeled by i) and the predicted value of the outcome variable for that observation within a given j$^{th}$ level 2 cluster (which is itself nested within a given k$^{th}$ level 3 cluster). $u_{0jk}$ is the difference between the overall value of the outcome variable for a given j$^{th}$ level 2 cluster (within a given k$^{th}$ level 3 cluster) and the overall prediction for the value of the outcome variable in said j$^{th}$ level 2 cluster. $\nu_{00k}$ is the difference between the overall value of the outcome variable for a given k$^{th}$ level 3 cluster and the overall prediction for the value of the outcome variable in said k$^{th}$ level 3 cluster. For example, examining student grades across multiple classes and multiple schools will yield a prediction for each student's grade. In standard regression the error term is simply the difference between a given student's actual grade and their predicted grade but in HLM there is a prediction for a given student who is in a given class which is in a given school. In this situation $\nu_{00k}$ is the difference between the overall grade for a given school and the overall predicted grade for that school, $u_{0jk}$ is the difference between the overall grade for a given class in a given school and the overall predicted grade for that class in that school, and $\varepsilon_{ijk}$ is the difference between a given student's actual grade and the predicted grade for that student in that class in that school.

When m is not 0 $u_{mjk}$ represents the difference between the regression coefficient for the m$^{th}$ level 1 predictor variable in the j$^{th}$ level 2 cluster within the k$^{th}$ level 3 cluster and the overall regression coefficient for the m$^{th}$ level 1 predictor variable in the k$^{th}$ level 3 cluster after accounting for all level 2 predictor variables. For example, say that in school 4 (so the 4$^{th}$ level 3 cluster) there is an overall regression coefficient (for some outcome variable, say grades) on the level 1 predictor variable TestScore (the score on some standardized test). There is also a regression coefficient on TestScore for class 6 in school 4 which may be different than the overall coefficient for school 4. The difference between these two coefficients would be $u_{364}$ (assuming TestScore is labeled as the third level 1 predictor variable) when the values of all class level variables are held constant.

When m is 0 but n is not 0 $\nu_{0nk}$ represents the difference between the regression coefficient for the n$^{th}$ level 2 predictor variable in the k$^{th}$ level 3 cluster and the overall regression coefficient for the n$^{th}$ level 2 predictor variable after accounting for all level 3 predictor variables. For example, say that there is an overall (across all classes and schools) regression coefficient on the level 2 predictor variable TeachYrExp (the teacher's temporal experience in years). There is also a regression coefficient on TeachYrExp for school 4 which may be different than the overall coefficient. The difference between these two coefficients would be $\nu_{014}$ (assuming TeachYrExp is labeled as the first level 2 predictor variable) when the values of all school level variables are held constant.

When neither m nor n are 0 $\nu_{mnk}$ represents the difference between the regression coefficient for the interaction term between the m$^{th}$ level 1 predictor variable and the n$^{th}$ level 2 predictor variable in the k$^{th}$ level 3 cluster and the overall regression coefficient for the n$^{th}$ level 2 predictor variable after accounting for all level 3 predictor variables. For example, say we posit that there is an interaction between the level 1 predictor variable TestScore and the level 2 predictor variable TeachYrExp (maybe teachers who have more years of experience better understand the tricks employed by the standardized test that is used to acquire students' TestScores). Then there is an overall (across all classes and schools) regression coefficient on this interaction term in the model. There is also a regression coefficient on this interaction term for school 4 which may be different than the overall coefficient.  The difference between these two coefficients would be $\nu_{314}$ when the values of all school level variables are held constant.

As with standard regression it is possible to find the variance of these error terms.  The variance in $\varepsilon_{ijk}$ (known as the residual variance and denoted $\sigma_{\varepsilon}{}^2$) represents the variation in the outcome variable within a given level 2 cluster (which again, is itself nested within a given level 3 cluster) after accounting for the effects of all predictor variables.  It is assumed that this variance is the same for all level 2 clusters unless stated and specified otherwise.  The variance in $u_{0jk}$ ($\sigma_{u0}{}^2$) represents the variation in the outcome variable between level 2 clusters within a given level 3 cluster after accounting for the effect of all level 2 predictor variables in the equation for the level 1 intercept ($\beta_{0jk}$).  The variance in $\nu_{00k}$ ($\sigma_{\nu 00}{}^2$) represents the variation in the outcome variable between level 3 clusters after accounting for the effect of all level 3 predictor variables in the equation for the level 2 intercept ($\gamma_{00k}$).

The variance in $u_{mjk}$ when m is not 0 ($\sigma_{um}{}^2$) represents the variation in the error term $u_{mjk}$ described previously.  Similarly, the variance in $\nu_{0nk}$ when n is not 0 and $\nu_{mnk}$ when neither m nor n are 0 ($\sigma_{\nu 0n}{}^2$ and $\sigma_{\nu mn}{}^2$ respectively) represent the variation in their respective error terms which were also described previously.  $\sigma_{\varepsilon}{}^2$, $\sigma_{u0}{}^2$, and $\sigma_{\nu 00}{}^2$ can be defined in this type of manner as well but it is informative to be more explicit about these which is why they were discussed separately.  This is because they can be used to determine intraclass correlation coefficients (ICCs) for level 2 and level 3.

$$\mathrm{ICC}_2 = \rho_2 = \frac{\sigma_{u0}{}^2}{\sigma_{\varepsilon}{}^2 + \sigma_{u0}{}^2 + \sigma_{\nu 00}{}^2}$$

$$\mathrm{ICC}_3 = \rho_3 = \frac{\sigma_{\nu 00}{}^2}{\sigma_{\varepsilon}{}^2 + \sigma_{u0}{}^2 + \sigma_{\nu 00}{}^2}$$

The ICC for a given level is the fraction or proportion (which can be converted to a percentage if desired) of variance in the outcome variable that is attributable to between cluster variations at that level [9, p. 34].  Therefore, $\mathrm{ICC}_2$ is the proportion of variance in the outcome variable that is due to variations between level 2 clusters in a given level 3 cluster while $\mathrm{ICC}_3$ is the proportion of variance in the outcome variable that is due to variations between level 3 clusters.  For example, the variance in student grades can be partially attributed to the variation between students in a given class in a given school ($\sigma_{\varepsilon}{}^2$), partially attributed to the variation between classes in a given school ($\sigma_{u0}{}^2$), and partially attributed to variations between schools ($\sigma_{\nu 00}{}^2$).  It is possible to define an $\mathrm{ICC}_1$ as the proportion of variance in the outcome variable that is due to variations between observations in a given level 2 cluster (which is itself nested within a given level 3 cluster) but this is generally not done both because it is often more important to know what proportion of the variance exists between level 2 clusters within a given level 3 cluster as well as between level 3 clusters than it is to know what proportion of the variance exists within a given level 2 cluster (within a given level 3 cluster) and because this would be redundant since $\mathrm{ICC}_1 + \mathrm{ICC}_2 + \mathrm{ICC}_3 = 1$.  The fact that variance can be parsed out and attributed to different levels in this way is yet another example of additional useful information that can be garnered from HLM which cannot be determined through standard regression techniques.  In HLM the first model that is typically analyzed is the Null Model which does not include any predictor variables since its objective is to figure out how much variance exists at different levels before any of this variance is explained (and therefore reduced) by incorporating predictor variables at one or more of levels.

Intercepts represent the average value of the outcome variable when all predictor variables take on a value of 0 (either for a given $j^{th}$ level 2 cluster within a given $k^{th}$ level 3 cluster, a given $k^{th}$ level 3 cluster, or overall if there is no subscript) making them intimately tied to error terms (with the "prediction" here being for a hypothetical observation where all predictor variables take on a value of 0). Slope coefficients (either for a given $j^{th}$ level 2 cluster within a given $k^{th}$ level 3 cluster, a given $k^{th}$ level 3 cluster, or overall if there is no subscript) on dummy predictor variables (predictor variables that are binary and have a value of either 0 or 1) represent the average amount by which the outcome variable is different for those who have the characteristic that is assigned a value of 1 compared to either those who have the characteristic assigned a value of 0 or the reference category when there is a set of complete and mutually exclusive dummy variables (a set of dummy variables where every observation is assigned a value of 1 for exactly one of these variables and 0 for all of the others) after controlling for all other predictor variables in the model. Slope coefficients (again, either for a given $j^{th}$ level 2 cluster within a given $k^{th}$ level 3 cluster, a given $k^{th}$ level 3 cluster, or overall if there is no subscript) on continuous predictor variables represent the average amount by which the outcome variable changes when the value of said variable increases by 1 unit after controlling for the effect of all other predictor variables in the model.

It is important to realize that these models can get quite complicated rather quickly (especially with more than two levels) so while the above descriptions were completely general in practice it is rare for these models to include all possible terms in their respective equations. Instead, theory and empirical evidence (frequently in the form of testing a series of increasingly more complex models while dropping non-significant terms along the way) along with philosophical notions like a desire to have the simplest possible model that still makes sufficiently good predictions are used to decide which terms to include and which ones to leave out. On a related note, standard regression techniques tend to fit data based on ordinary least squares (OLS) procedures where the best fit line is determined by minimizing the sum of the squares of the residuals (error terms). HLM, however, typically uses maximum likelihood estimation (MLE) where the best fit line is determined by maximizing some likelihood (probability) function (the details of which are beyond the scope of this paper). This method can also be used to do standard regression, though it is not possible to use OLS in HLM without substantial modifications because there are different types of error terms at different levels and how to appropriately weight and properly use each of them is not well-defined unless explicitly specified.

# Site and Sample

The observations used in this study (the sample) were drawn from five years (2012 – 2016) of data on student Grades and characteristics in a series of three introductory physics courses designed for (and taken primarily by) bioscience majors at a large, public, R1 university that is on the quarter system (the University from here on out). These courses are based on active learning techniques where during the regular academic year (Fall, Winter, and Spring quarters) students spend five hours per week in discussion-lab (DL) where they engage in activities in groups of about five students each and synthesize the material as a whole class, both with guidance from a teaching assistant (TA). An additional 1.5 hours per week are spent in lecture. During the regular academic year lecturers teach two lectures, of about 165 students each, back-to-back for a total of about 330 students who are broken into DLs of approximately 30 students each (though sometimes overflow classrooms are used which hold approximately 20 students each). Approximately half of the students in most DLs come from the first of the aforementioned lectures while the other half come from the second (so DLs usually consist of a mixture of students from two related, but distinct, lectures).

# Data

The data used in this study was acquired by a professor at the University for the purposes of education research. This study will examine student understanding of physics by clustering students into three nested levels; individual students at level 1 (really observations but this essentially amounts to students), DLs at level 2, and Lectures at level 3, such that students (observations) are nested within DLs and DLs are nested within Lectures. For all of the models used in this study i indexes individual students (or really, observations since there are students who are associated with more than one observation), j indexes DLs, and k indexes Lectures. In this context a Lecture refers to both lecture sections taught back-to-back by the same lecturer during the regular academic year since they tend to be quite similar and it would be difficult to deal with the DL nesting structure if they were treated as distinct. The different lecture start times within the same "Lecture" are taken into consideration as discussed below.

Data taken from graduate students was excluded from the analysis because graduate students are likely to be qualitatively different in a multitude of ways than the rest of the students in the sample and there were only 28 observations associated with graduate students anyway. Observations associated with a course drop were also excluded from the analysis because dropping a course (which initially corresponded to a final numerical grade of 0) is a different outcome than failing it (which also corresponds to a final numerical grade of 0) and even beyond this, dropping a course is a fundamentally distinct outcome from receiving a numerical grade (so it would not make sense to simply recode these data points as having some Grade other than 0). Finally, for the purposes of this analysis observations associated with summer courses (in both Summer Session I and Summer Session II) were excluded since the types of students who take summer courses are often different than those who take these same courses during the regular academic year and DL (as well as lecture) start times can be different over the summer than they are during the regular academic year (plus the Lecture structure is also different during the summer than the back-to-back format mentioned above since during the summer lecturers typically only teach one lecture).

# Outcome Variable

The outcome variable in this study is Grade which represents the final numerical (on a 4.00 scale) course grade for a given student in a given class (DL plus Lecture). Therefore, this outcome variable is assumed to be continuous and exists at level 1 of the clustering (the individual student level in this case) which makes it relatively straightforward to deal with using standard multilevel modeling techniques. The numerical Grade associated with an A+ letter grade was changed from the original value of 4.00 to a value of 4.33 in order to match the association between other letter grades and their corresponding numerical grades and because oftentimes students still try to get an A+ even when they know it will not affect their GPA any differently than an A would. This also made the distribution of final numerical Grades closer to a Normal Distribution, though there is still some noticeable skew as well as a ceiling effect and a disproportionately high number of 0s (Fs).

The assumption of continuity (which also applies to the variables GPA, Units, and LecSize in this study) has two major components. First off, it means that the number of possible values is fairly large but what “fairly large” entails is not well defined because 13 values (the number of standard letter grades including “A+”) and 101 values (the number of integers on a 100 point scale) are both discrete and, strictly speaking, neither of them is truly continuous. The only difference between the two is how many distinct values (grades) are allowed and the amount by which consecutive values are separated and yet, while the accuracy of treating letter grades as continuous is sometimes disputed the continuity of 100 point scales is rarely questioned. The second component is directly tied to this question of value separation and whether the difference between consecutive values is meaningful. For ordinal variables this is not assumed to be true and consecutive values simply represent an ordered ranking without any meaning attached to the degree of separation. For continuous variables it is assumed that the difference between consecutive values has the same meaning anywhere along the spectrum. For instance, treating Grade as continuous here means that the difference between a C- (1.67) and a C (2.00) is assumed to be the same as the difference between a B (3.00) and a B+ (3.33) which is not necessarily the case. However, the same would be true on a 100 point scale where it is not necessarily the case that the difference between say, a 60 and a 61, is the same as the difference between an 84 and an 85 but in both cases this assumption needs to hold in order for numerical grades to be treated as continuous variables and in both cases it is reasonable to assume that it does hold to good approximation (and if not there would be a lot of other philosophical and pedogeological problems that would need to be contended with).

On top of these potential issues (which have hopefully been addressed to the reader’s satisfaction), curves are sometimes implemented when it comes to grading introductory physics courses in which case one could argue that these grades are a ranking system within a given class (an ordinal level of measurement) but do not present an interval level of knowledge since the difference between consecutive grades at one point along the grade spectrum does not necessarily mean the same thing as the difference between consecutive grades at a different point along this spectrum. Another way that grades are sometimes manipulated which may raise concerns around the degree to which they can be treated as continuous is shifting the thresholds or cutoffs for obtaining a given letter grade (a different type of correction for relative difficulty which is not the same as a curve). However, as far as the author is aware the courses involved in this study are rarely curved and try to avoid shifting the thresholds for obtaining a given letter grade (compared to some conventional standard).

# Grades in the Context of Measurement Theory

Considering the fact that the outcome variable here is overall course grades for individual students it is clear that the nature, meaning, precision, and accuracy of individual grades is foundational to this study. In the context of our current society grades are relevant in-and-of themselves because they are a common feature of most students' educational experiences and are used as sorting or ranking mechanisms to determine who qualifies for certain benefits [10]. Essentially, reviewers often use grades to help determine which students (or former students) have access to resources and opportunities such as jobs or additional levels of schooling (like internships, medical school, and graduate school). This is the primary reason that most students tend to be concerned about their grades and the courses involved in this study are no exception. These uses therefore lead directly to questions around the meaning, legitimacy, and interpretation of grades which is itself a question of their validity based on the most widely accepted definition of the term, though it should be noted that there is still no fully agreed upon definition [11, p. 255].

One frequently held belief is that grades are "what students 'earn' for their achievement" [10]. This assertion assumes that grades are based solely on students' understanding of the relevant subject matter which, if true, would be useful here since the goal of this study was to explain and predict students' understanding of physics principles, the underlying construct (latent trait) of interest in this case. This is an important construct both because people tend to be curious about the fundamental laws of nature and studying physics is a good way of learning about such things and because the problem solving skills required to analyze physics problems are applicable to other scenarios as well and the concepts themselves are useful in a variety of ways from everyday situations to more career specific inquiries. In addition to students who take physics courses being concerned about their physics grades for the reasons described above many of them probably have at least some curiosity about the laws of physics (the construct of interest here) both in general and in terms of how physics relates to their chosen fields and majors (so for the most part, biological applications of physics when it comes to the courses involved in this study). Since it is not possible to directly observe someone's physics understanding this study took final course grades to be a test (in a generalized sense) that measures such things [11, p. 3].

However, despite some commonly held beliefs there are a range of problems with the assumption that grades are exclusively, or even largely, a measure of academic achievement. For one, in K-12 schooling numerous studies of concurrent and predictive (criterion-related or using more contemporary terminology; evidence based on relations to other variables) validity have been conducted to examine the relationship between grades (sometimes overall grades like GPAs and sometimes grades in specific classes, like math) and outcomes on so-called "achievement" or "intelligence" tests. These studies have consistently demonstrated (even as the composition of such tests and the educational system more broadly have gone through substantial changes over the years) a moderate relationship between the two implying that grades are related to achievement (or "intelligence") as defined by such tests (the criterion in this case) in a significant but modest way [10]. It should be noted that these sorts of studies assume that such tests actually measure achievement (or "intelligence") and going even deeper, that the constructs of "achievement" and "intelligence" have well-defined meanings (i.e. that these tests are themselves valid). While university courses are obviously different than K-12 schooling it is likely that similar relationships exist there as well, though it would be beneficial to make this determination for certain through empirical studies. For instance, one possible format for studies of this type could be to correlate grades in introductory college physics courses that teach Newton's Laws with scores on the Force Concept Inventory (FCI).

Building off of the above, it eventually became evident through empirical studies of grading practices that K-12 report card grades are multidimensional measures of a variety of cognitive and noncognitive factors that both assess student learning and motivate it based on what teachers value in student work. These often include such things as achievement, substantive engagement, persistence, improvement, and even consequences of grades on student success and feelings about their competence [10]. These studies align well with teachers' perceptions of their own grading practices as determined by surveys and interviews where teachers brought up the inclusion of noncognitive factors in the grades they assign along with many teachers expressing a desire to grade fairly which to them meant using multiple sources, incorporating effort, and making grading policies clear to students. Context and professional judgment is sometimes included as well rather than relying solely on a grading algorithm. However, teachers' beliefs and values determine the purpose and extent of the impact that factors which are not directly tied to achievement have on grades and these vary between teachers on both an individual and group level, sometimes within the same school and possibly even between students who have the same teacher as a result of differing contexts [10].

On the group level, modern elementary school teachers largely think of grades as being more about communication with students and parents while secondary school teachers think of them more in terms of classroom management and place a higher value on exams [10]. While it is likely that there exists a range of grading schemes in higher education, especially when grading standards are considered to be a matter of academic freedom in the U.S., one might suspect that these trends continue and that college instructors incorporate a variety of factors into the grades they assign but emphasize the exam-based achievement side of grades more than K-12 teachers do. Once again, it would be beneficial for additional studies to be conducted so a more definitive determination can be made but in the author's experience this could certainly be argued in the case of many introductory college physics courses, including those involved in this study, where timed exams and quizzes are the primary basis for course grades while participation (nonachievement) and homework (a combination of achievement and nonachievement since students can work together on it) play a significant but rather limited role. It would also be helpful for more through evaluations of grades in K-12 schooling as well as higher education to be done using factor analysis (both exploratory and confirmatory based on previous research) in order to better understand and account for the different components that go into student grades both in general and in particular situations.

One possible way to address discrepancies between the relative weights given by individual teachers to different components of grades would be to standardize these weights by requiring the proportion of grades attributable to each component be the same across all students or at least all of those at the same grade level. This may not go over well with many teachers though and academic freedom combined with grading autonomy would make it difficult to enforce. A related but distinct possibility would be to give separate scores for different components as is the case with standards based grading. This would recognize the importance of different attributes but would systematically and consistently distinguish between them [10]. It would even be possible to include a section on context or professional judgement. However, these sorts of suggestions ignore the fact that some teachers consider behavior that promotes academic achievement to be part of academic achievement and therefore, may discount any attempts to separate such things on principle [10]. Furthermore, even though differential impact is not considered "bias" on its own in measurement theory, many teachers are rightly concerned about the negative material consequences that low grades and their intersection with other student characteristics (like race, gender, and socioeconomic status) can lead to under the current system and may be reluctant to separate such things as a result knowing that low scores in certain areas will likely reproduce institutional violence even when reported in conjunction with high scores in other areas [11, p. 478]. The only way to remedy this concern would be to dismantle capitalism and other forms of oppression by restructuring the way society functions such that people are no longer punished simply for having low academic achievement scores.

One last point that should be brought up in this discussion of grades' criterion-related validity is that in K-12 schooling grades are known to predict drop-out rates and other measures of success and failure in subsequent levels of schooling [10]. However, one could argue that this is circular reasoning since it is unsurprising that measures of "success" in schooling relate to other measures of "success" in schooling regardless of what any of these measures actually get at. It is interesting to note, however, that standardized achievement tests do not predict such things nearly as well which raises questions about the meaning, interpretation, and uses of both standardized achievement tests and grades [10].

Finally, it is important to consider the content validity of grades' achievement component. More specifically, since grades for the courses involved in this study are primarily based on quizzes and exams it could be argued that they are largely reflective of student achievement but even if true this does not necessarily guarantee that they measure the types of achievement that both physics instructors and the aforementioned reviewers believe they do. In particular, not only are quizzes and exams just one potential assessment of achievement among many but these quizzes and exams tend to be timed and necessitate short, written responses (as is typical with introductory physics problems). Considering the construct of interest for this study as well as instructors and, presumably, application reviewers is students' understanding of physics principles this format impacts student grades by both introducing some degree of construct-irrelevant variance and causing some degree of construct underrepresentation [11, p. 261].

A few possible sources of construct-irrelevant variance here are speediness, ability to perform under pressure, reading comprehension, writing abilities, and handwriting. These factors lead to some variance in quiz and exam scores that is independent from the construct of interest and may even introduce bias into these scores as well as course grades that are heavily dependent on them. This potential bias is because of the different ways that such things affect various groups [11, p. 479]. For instance, if physics quizzes and exams require a certain level of reading comprehension or writing ability then scores on them are likely to be lower for non-native English speakers than for their native English speaking counterparts because of language barriers that have nothing to do with physics knowledge or understanding. Similarly, timed exams are known to lower scores for women and girls disproportionately more than they lower scores for men and boys. Much (though not necessarily all) bias of this sort could be adequately addressed by quizzes and exams that follow principles of universal design [11, p. 503]. Construct underrepresentation would come from restricting the types of achievement that grades in these courses capture by largely leaving out such things as lab skills, oral processing, and being able to productively expand upon others' ideas while focusing almost exclusively on direct knowledge and calculational abilities. This too could potentially bias course grades because of differences in what various groups are socialized to value such that certain groups may be better at the types of skills that are being tested while others may be better at those that are not. It is also quite possible that students end up focusing on the types of behaviors and skills that will yield the highest possible course grades given the primary uses of grades described above in which case this construct-irrelevant variance and construct underrepresentation could easily have the effect of shaping what students prioritize when trying to learn the material. It would therefore be beneficial for future research to determine the types of skills that are measured by conventional physics exams and quizzes (perhaps using factor analysis) along with how this impacts students' study habits. It would then be up to the physics education community to decide if these are desirable skills to measure as well as whether they are the only skills that should be measured and to adjust accordingly.

Going even deeper, not all aspects of the material that is covered in introductory physics courses, such as those involved in this study, is represented on the quizzes or exams for any given course and even when it comes to the material that is represented the breakdown of how much each topic contributes to students' grades can vary widely. It would be nice if physics instructors could come to some sort of consensus on which topics should be covered in introductory courses, their breakdown in terms of grading, and the breakdown of difficulty levels within each topic and then largely stick to this structure. However, since this is unlikely to occur each individual instructor should at least take the time to think through their own feelings on these matters and be explicit with themselves, their peers, and their students about such things through personal specification tables [11, p. 269]. The need for content experts to evaluate quiz and exam problems is less clear since it would not be reasonable to require every quiz or exam problem to go through this process and most physics instructors would consider themselves to be content experts when it comes to introductory physics principles anyway. There is also some dispute over who would qualify as an expert in this regard and whether, for instance, professional physicists who are not teaching faculty would be effective at making these sorts of determinations [12]. However, there might be some need for the courses involved in this study to have their quizzes and exams, or at least a range of sample problems for instructors to base quizzes and exams off of, to go through some sort of expert evaluation since these courses are not only taught with a different format than traditional lecture-based physics courses but they also use a different curriculum that physics instructors who have been trained through more conventional means may not be familiar with. Obviously, none of these suggestions can be enforced is most cases because of academic freedom and grading autonomy but individual instructors are still able to adhere to best practices when creating assessments (like quizzes and exams) and should strive to do so.

Another related issue is the potential mismatch between what is taught in DL and what appears on quizzes and exams since DL is taught by a TA while quizzes and exams for the courses involved in this study are usually designed and implemented by the instructor and given in lecture. The lectures for these courses are supposed to coincide with what is taught in DL but this is not always the case both in terms of the standard DL curriculum (which is rather rigid) and in terms of how DL is actually taught in practice which can vary between DLs based on TA style, potential modifications, etc. This is an extremely relevant question for this study since the goal here is to evaluate the relationship between course grades and classroom characteristics, especially class (DL) size, which requires course grades to be a relatively strong measure of what is taught in DL. For the purposes of this study it was assumed that despite all of their flaws, student grades in the courses involved here at least measure the same overall content that is taught in DL but it would be worth conducting a rigorous analysis of this relationship in the future. Similarly, it was assumed that students are malleable and that their learning depends on the classroom environment and instructional guidance since if this was not the case then classroom characteristics would clearly have no impact on their understanding of physics. This too is a question that could use further study.

While the validity of grades is important both in general and as a framework to help guide the meaning and limitations of this study, evaluating the merits of grades does not end there. Beyond the extent to which the intended uses of grades match what they actually measure, another concern that anyone who uses grades to judge student performance, from researchers to instructors to application reviewers, should have is simply the degree to which grades consistently measure something, be it academic achievement or a broader multidimensional mix of traits. This is where reliability comes in. In general, it is possible that approaches to grading which separate out different components of grades, such as standards based grading, could make grades more reliable because each grade would then measure a single construct. However, it has been found that even under more traditional models of grading as assignments are aggregated up to course grades the reliability of grades tends to increase regardless of what it is that they are measuring [10]. This means that in a typical scenario a greater proportion of students' true scores (to use terminology from Classical Test Theory) on whatever construct(s) a given course's grade is measuring is represented by their overall grade in that course than by their grade on any given assignment which makes sense because one would expect the effect of random error to decrease as more assignments are taken into account [11, p. 161-162]. This implies that overall course grades are likely to be a better proxy for students' understanding of physics concepts in the courses involved here than grades on individual assessments would, assuming that both overall course grades and individual assessment grades overwhelmingly measure physics understanding rather than a multitude of constructs which is likely the case for the courses involved in this study as has already been discussed. It is true, however, that even in these courses overall grades likely measure other constructs to some degree whereas quiz and exam grades likely do so to a lesser extent so it is still possible that grades on certain individual assessments might be better indicators of students' physics understanding than overall course grades and this could be a good question for future research. It is for this reason as well as the fact that overall grades for the courses involved in this study are composed primarily of grades on individual quizzes and exams that the reliability of individual assessments is a vital part of any discussion around the reliability of course grades for the purposes of this study and so that is where this section now turns.

For any given individual assessment there are a variety of reliability concerns to consider. First off, there are questions of internal reliability (whether or not different items are measuring the same construct(s)) and alternate forms reliability (whether or not different forms are measuring the same construct(s)) if different forms of the assessment are given. These two types of reliability are difficult to study in general since they depend on the nature of the assignment in question which can vary tremendously across instructors, classes, schools, etc. because of the large degree of autonomy that teachers usually have in coming up with assignments both in K-12 schooling and especially in university settings where this is often a matter of academic freedom. However, it is once again the case that individual instructors, including those who teach the courses involved in this study, can still adhere to best practices and should strive to do so. More specifically, to address these concerns they could potentially conduct their own reliability studies by, for instance, finding coefficient alpha or correlating alternate forms for a typical exam (though these analyses may be rather cumbersome) even if it is not required that they do so. Furthermore, administrators should encourage them to do so both informally and through direct incentives like considering these practices when conducting teacher evaluations and making decisions around promotions. A similar line of reasoning can be applied to item bias on quizzes and exams where it is difficult to study such bias in general because it is unique to individual problems (though one could conduct general studies of item bias on common types of problems which is certainly something that exists in introductory physics courses) but it would be good practice for individual instructors to conduct their own analyses of item bias and discard items accordingly and they should be encouraged to do so [11, p. 483-499]. However, regardless of suggestions for future practices there is no way to tell what the internal or alternate forms reliability of the quizzes or exams used during the courses involved in this study were nor is there any way to tell how biased they were as a result of either individual item bias or more holistic factors like those described earlier. And yet, such things certainly affect the reliability and validity of course grades and thus, the statistical power of regression analyses that include course grades as a variable, so these unknown pieces of information present a definite limitation to this study that cannot be ignored.

Another important aspect of reliability to consider is inter-rater reliability or the consistency with which different graders grade the same assignment. This aspect of reliability is possible to study in general and many such studies (focusing on K-12 teachers, though their conclusions likely extend to university instructors as well given the individualized nature of grading preferences) have been conducted, particularly during the early 1900s [10]. These studies largely showed a significant degree of variation between teachers (about 5 points on a 100 point scale), though a few studies disagreed with this conclusion. The primary sources of variation were found to be an inability to distinguish between assignments of similar "merit" (which can be conceptualized as random error), differences between teachers' grading standards, and differences between the relative weights that teachers assigned to different aspects of an assignment [10]. It is not much of a stretch to imagine that bias could be a relevant factor here as well (even if it is not one that many prominent academics thought about during the early $20^{th}$ century) since grader biases have the potential to show up in the grades that they assign and different graders have different types and levels of bias. This variability in the grades that different teachers involved in these studies would give to the same assignment eventually led to the development and implementation of letter grading in an attempt to reduce the effect of rater uncertainty on grades which bolsters the argument put forward in the previous section that letter grades are an appropriate outcome variable to use in this study [10].

However, the methodologies used in these early studies of inter-rater reliability had their flaws. For example, teachers were often sent assignments to grade without specific grading criteria [10]. Because of this, some of the uncertainties in grading that were identified by these studies could be reduced through a range of improvements from using better grading criteria that incorporates student input to more collaboration among teachers when it comes to grading practices [10]. Wider adoption of standards based grading as discussed earlier could also help by parsing out the different components of grades and effectively standardizing the aforementioned weights [10] as would the formation and implementation of better grading criteria in the form of more rigorous and standardized rubrics [13] or grading by category [14]. More research would need to be done before making any assertions about the reliability of these techniques but they seem to have at least some promise of generalizability. Either way though the courses involved in this study already use rubrics and grade by category and each problem is usually graded by a single TA anyway so while there is always room for improvement inter-rater reliability is probably not a major limitation for this study.

One last way of potentially increasing the inter-rater reliability of individual assessments, at least in physics courses, would be to use assessments that require less subjective grading by, for example, employing multiple choice questions that approximate certain aspects of free-response questions. While the two will never be equivalent there are some preliminary results suggesting that it is possible for multiple choice questions to mimic their free-response counterparts under the right circumstances [15]. This is especially true if incorrect answers on the multiple choice version conform to common mistakes that students often make in free-response form (the fact that common mistakes can be categorized in this way is the basis for categorical grading systems as mentioned above) and different levels of partial credit are given to incorrect multiple choice answers in a similar manner to how partial credit would be assigned to similarly incorrect free-response answers [15]. This leads to a much longer discussion about the relative merits of different assessment formats but purely from the perspective of measurement theory it would require, at minimum, much more extensive research on alternate forms reliability between free response problems and exams and their multiple choice counterparts by developing a large question bank which includes both versions of each question; administering them to a large, representative sample; and correlating scores on the two versions. Assessments formed from these questions would also have to be evaluated for internal reliability and free response problems would have to be checked for inter-rater reliability during the research portion of such a program in order to make sure that all of these items are measuring the same construct and that when the multiple choice version of a question is compared to the free response version there is a well-defined and agreed upon free response score to use as a reference point.

Taken together all of the above implies that course grades are a reflection of a range of important traits that the application reviewers who use them most frequently are likely to be interested in (provided they actually care to evaluate applicable constructs rather than simply reproduce society's hierarchies, intentionally or not) from academic achievement to communication skills to team work to effort and perseverance. However, the degree to which different characteristics contribute to grades can differ quite substantially and a lot of people, possibly even including many application reviewers, do not realize any of this and instead believe that grades are purely a reflection of academic achievement which at the moment is both not true for the most part and not necessarily desirable given the importance of various other attributes. It must also be acknowledged that even the achievement component of grades does not always reflect the full range of academic ability that one might expect it to and plenty of people are likely unaware of this as well. False beliefs about the nature and interpretation of grades can therefore impact the decisions that are made based on them in a way that does not properly reflect their true meaning or appropriate uses which is something that should be addressed by administrators, managers, politicians, and others who have power over relevant policies under the current system. In some cases though grades do largely reflect academic ability and not much else and it would seem as if overall grades for the courses involved in this study are among them meaning these grades are a fairly good proxy for physics understanding (whether or not the types of physics understanding that they reflect correspond to the type of material that is taught in DL which is a separate question) and are therefore a fairly good outcome variable to use in this study. They are clearly not perfect though and there are still some problems with using them in this way that future research will hopefully shed more light on. Perhaps it will turn out that grades on certain individual assessments, like final exams, are better for this purpose or that something entirely different from grades, such as scores on the FCI or an analogous assessment, would be best but in the meantime overall course grades appears to be a relatively good approximation.

# Predictor Variables

## Level 1

Male is a dummy variable representing the binary sex that students were identified with. It is 1 for male-identified students and 0 for female-identified students since there are more females than males in the sample. LecStart is also a dummy variable and represents the start time of the lecture that students attended. It is 0 for the earlier (7:30am) lecture and 1 for the later (9:00am) lecture during the regular academic year.

Units and GPA are both continuous variables which represent the number of units that students had gotten credit for and their GPA at the University prior to the quarter in which data was taken, respectively. Both of these are relatively Normal but have some skew and there is a major ceiling effect for GPA and floor effect for Units. Note that GPA was not recorded in situations where Units was less than 12 (one quarter’s worth for a full-time student). This means that the conclusions drawn from this study will not necessarily apply to first quarter freshmen or transfer students (groups whose members may or may not have taken one of these courses during their first quarter at the University) because students from these groups are disproportionately excluded from the data used in the analysis. This is in addition to the fact that the conclusions drawn from this study will not apply to the few graduate students who take these courses or students who take them over the summer since students in these groups were also systematically excluded from the data used in the analysis.

## Level 2

DLSize is a continuous variable which represents the number of students in a given DL and is fairly Normal but has some noticeable skew. Because it is possible that the effect of DL size would be non-linear (for instance, if students learn more in classrooms with around 18 students and learn less in classrooms with both greater and fewer numbers of students than this) and because DLSize is strongly peaked around 30 students meaning relatively small DL size differences in this range could dominate the statistics when it is not expected that there would be any significant differences between DLs of say, 28 and 31 students (in other words, DLSize is not expected to be continuous in the second sense described above), DLSize was converted to a set of seven dummy variables. These are RlySm (under 9), Sm (9-14), Lit (15 -20), Med (21-26), Stand (27-32), Lg (33-38), and RlyLg (over 38) that were used in the analysis in place of DLSize. Stand represents the standard range of sizes for DLs in EPS rooms (see below for more on different DL rooms) and is the reference category here while Lit represents a range of sizes that fall around those which have been found to be ideal in the literature. RlySm (Really Small), Sm (Small), Med (Medium), Lg (Large), and RlyLg (Really Large) are fairly self-explanatory. In EPS DL rooms there are six tables (and therefore, six groups) so it made sense for each DL size dummy variable to incorporate six values of DLSize. This is also similar to the ranges used for class size dummy variables in other class size studies.

DLStart is a continuous variable which represents the start time of said DL in hours (where minutes were converted to fractional hours given in decimal format and rounded to two places) based on a 24-hour cycle. There are only five possible DL start times during the regular academic year. These are 8:00am (8), 10:30am (10.5), 2:10pm (14.17), 4:40pm (16.67), and 7:10pm (19.17). Because the number of DL start times during the regular academic year is relatively small and one might suspect that the effect of DL start time on Grade would not necessarily be linear (for example, maybe really early and really late start times have a similar effect which is different than the effect of midday start times) it is desirable to treat DL start times as a set of dummy variables in the analysis rather than a single continuous variable. However, because each DL meets two days per week during the regular academic year there are instances where a DL will meet at one of the five standard start times on the first day and a different one on the second day. In principle any combination of two DL start times is possible but the ones that actually occurred in the data were DL start times of 10:30am and 2:10pm, 8:00am and 4:40pm, 10:30am and 4:40pm, and 2:10pm and 4:40pm. In these cases, the two different DL start times (in hours) were averaged to produce that DL's recorded start time with the exception of 8:00am and 4:40pm which was recorded as 12.34 instead of 12.33 to distinguish it from the 10:30am and 2:10pm situation. The existence of this type of scheduling which may have effects beyond that of the average start time is yet another reason to treat DL start times as a set of dummy variables rather than a single continuous one. Each of these situations was treated as its own dummy variable in the analysis (DL1233, DL1234, DL1358, and DL1542 respectively) along with dummy variables for each of the standard DL start times (DL8, DL105, DL1417, DL1667, and DL1917 respectively) with DL1417 (2:10pm) serving as the reference category.

ROS is a dummy variable that is 1 for observations associated with a DL in ROS, a particular building whose rooms have particular layouts, and 0 for those associated with a DL in EPS which is a different building whose rooms have a different layout than those in ROS and are designed to accommodate DLs for the courses involved in this study (most DLs for these courses are held in such rooms which is why EPS rooms were coded as 0 here). ROS rooms are used as overflow rooms when needed. A few DLs met in an EPS room during one day per week and a ROS room on the other day. These were accounted for using another dummy variable (MultiRm) which was 1 for DLs that were like this and 0 for those that were not. These DLs were assigned a ROS value of 0. When doing the statistical analysis it turned out that MultiRm is essentially a combination of DL1358 and DL1542 (all DLs with either of these start times had one DL per week in a ROS room and the other one in an EPS room and no DLs with any other start times had such an arrangement) so it was dropped from the statistical analysis but conceptually it is still useful to know that these types of DLs exist in the data and are effectively accounted for through the variables DL1358 and DL1542.

Fall, Winter, and Spring are a set of dummy variables which represent the quarter in which a given DL was held. They are 1 for DLs that occurred during that quarter and 0 for all others with the reference category being Fall. Similarly, SevA, SevB, and SevC are a set of dummy variables which represent the course a given DL was part of. They are 1 for DLs that were part of that course and 0 for all others with the reference category being SevA.

Finally, the DL-mean of each level 1 variable was used as a continuous predictor variable at level 2. In particular, Mean_GPA and Mean_Units were used because in an active learning setting the intention is for students to teach each other and thus a DL with more overall background knowledge on the part of students (again assuming that course grades accurately reflect students' understanding and knowledge of the underlying material) might be expected to positively influence the understanding of new material that individual students in that DL acquire. For a variety of social reasons, the average number of women in a classroom, which will likely be strongly related to Mean_Male in a given DL, can also have a significant effect on individual student learning in that classroom.

## Level 3

LecSize is a continuous variable which represents the number of students in a given Lecture.

In all of the models beyond the Null Model (see below for more about the different models used in this study) there were 20,280 total observations clustered into 787 DLs and 75 Lectures after observations with missing values for any one or more of the variables used in the analysis were removed. These missing observations are assumed to be randomly distributed within the sample. See Table 1 for summary statistics of the sample data on the variables used in this study.

| Variable | Obs | Mean | Std. Dev. | Min | Max |
|---|---|---|---|---|---|
| Grade | 21,829 | 2.919337 | .9313294 | 0 | 4.33 |
| GPA | 20,281 | 3.100617 | .4564486 | 1.166667 | 4 |
| Units | 21,829 | 69.15997 | 33.96601 | 0 | 219 |
| Male | 21,650 | .3653118 | .4815286 | 0 | 1 |
| LecStart | 21,829 | .5223327 | .4995124 | 0 | 1 |
| DLSize | 21,829 | 28.8989 | 3.920994 | 8 | 39 |
| RlySm | 21,829 | .0003665 | .0191407 | 0 | 1 |
| Sm | 21,829 | .0099867 | .0994356 | 0 | 1 |
| Lit | 21,829 | .0347703 | .1832016 | 0 | 1 |
| Med | 21,829 | .1495259 | .3566142 | 0 | 1 |
| Stand | 21,829 | .6836319 | .4650691 | 0 | 1 |
| Lg | 21,829 | .1199322 | .3248897 | 0 | 1 |
| RlyLg | 21,829 | .0017866 | .0422316 | 0 | 1 |
| DLStart | 21,829 | 13.60968 | 3.757423 | 8 | 19.17 |
| DL8 | 21,829 | .1593293 | .3659913 | 0 | 1 |
| DL105 | 21,829 | .2086216 | .4063326 | 0 | 1 |
| DL1417 | 21,829 | .19836 | .3987738 | 0 | 1 |
| DL1667 | 21,829 | .2001924 | .4001534 | 0 | 1 |
| DL1917 | 21,829 | .1628568 | .3692434 | 0 | 1 |
| DL1233 | 21,829 | .0673416 | .2506184 | 0 | 1 |
| DL1234 | 21,829 | .0012369 | .0351484 | 0 | 1 |
| DL1358 | 21,829 | .0015118 | .0388527 | 0 | 1 |
| DL1542 | 21,829 | .0005497 | .0234404 | 0 | 1 |
| ROS | 21,829 | .0160795 | .1257843 | 0 | 1 |
| MultiRm | 21,829 | .0020615 | .0453577 | 0 | 1 |
| Fall | 21,829 | .3109167 | .4628794 | 0 | 1 |
| Winter | 21,829 | .3496266 | .4768629 | 0 | 1 |
| Spring | 21,829 | .3394567 | .4735358 | 0 | 1 |
| SevA | 21,829 | .3804572 | .4855104 | 0 | 1 |
| SevB | 21,829 | .3386779 | .473271 | 0 | 1 |
| SevC | 21,829 | .2808649 | .4494319 | 0 | 1 |
| Mean_GPA | 21,829 | 3.102115 | .1163914 | 2.611243 | 3.45866 |
| Mean_Units | 21,829 | 69.15997 | 14.10221 | 35.58621 | 112.2424 |
| Mean_Male | 21,829 | .3649083 | .1097605 | 0 | .875 |
| Mean_LecSt~t | 21,829 | .5223327 | .0794057 | 0 | 1 |
| LecSize | 21,829 | 300.2445 | 46.87964 | 152 | 367 |

Table 1: Summary Statistics

# Analysis

This study was conducted in five stages. In the first stage a Null Model was specified and fit in order to determine the amount of variance in Grade between observations as well as how much of this variance lies within a given DL (which is itself nested within a given Lecture), between DLs but within a given Lecture, and between Lectures.

## Null Model

Level 1:

$$\text{Grade}_{ijk} = \beta_{0jk} + \varepsilon_{ijk}$$

Level 2:

$$\beta_{0jk} = \gamma_{00k} + u_{0jk}$$

Level 3:

$$\gamma_{00k} = \pi_{000} + \nu_{00k}$$

In the second stage an Individual Model was specified and fit which included all level 1 predictor variables but fixed their slope (regression) coefficients, meaning their slopes were not allowed to vary between DLs or Lectures, in order to control for their effects and determine what impact they have on Grade (as well as the variation between grades) when no level 2 or level 3 predictor variables are included.

## Individual Model

Level 1:

$$\mathrm{Grade}_{ijk} = \beta_{0jk} + \beta_{1jk} * \mathrm{Male}_{ijk} + \beta_{2jk} * \mathrm{LecStart}_{ijk} + \beta_{3jk} * \mathrm{Units}_{ijk} + \beta_{4jk} * \mathrm{GPA}_{ijk} + \varepsilon_{ijk}$$

Level 2:

$$\beta_{0jk} = \gamma_{00k} + u_{0jk}$$

$$\beta_{mjk} = \gamma_{m0k} \;\; \forall m \in \{1, \ldots, 4\}$$

Level 3:

$$\gamma_{00k} = \pi_{000} + \nu_{00k}$$

$$\gamma_{m0k} = \pi_{m00} \;\; \forall m \in \{1, \ldots, 4\}$$

In the next stage a DL Model was specified and fit which included all level 2 predictor variables in order to determine their effects on Grade (as well as the variation between grades) when none of the slope coefficients on the level 1 predictor variables were allowed to vary between DLs or Lectures, none of the slope coefficients on the level 2 predictor variables were allowed to vary between Lectures, and no level 3 predictor variables were included.

## DL Model

Level 1:

$$\mathrm{Grade}_{ijk} = \beta_{0jk} + \beta_{1jk} * \mathrm{Male}_{ijk} + \beta_{2jk} * \mathrm{LecStart}_{ijk} + \beta_{3jk} * \mathrm{Units}_{ijk} + \beta_{4jk} * \mathrm{GPA}_{ijk} + \varepsilon_{ij}$$

Level 2:

$$\begin{aligned}\beta_{0jk} = \gamma_{00k} + \sum_{n=1}^{6} \gamma_{0nk} * [\mathrm{DummyDLSize}]_{jk} + \sum_{n=7}^{14} \gamma_{0nk} * \mathrm{DL}[\mathrm{Time}]_{njk} + \gamma_{015k} * \mathrm{ROS}_{jk} + \gamma_{016k} \\ * \mathrm{Winter}_{jk} + \gamma_{017k} * \mathrm{Spring}_{jk} + \gamma_{018k} * \mathrm{SevB}_{jk} + \gamma_{019k} * \mathrm{SevC}_{jk} + \gamma_{020k} \\ * \mathrm{Mean_Male}_{jk} + \gamma_{021k} * \mathrm{Mean_LecStart}_{jk} + \gamma_{022k} * \mathrm{Mean_Units}_{jk} + \gamma_{023k} \\ * \mathrm{Mean_GPA}_{jk} + u_{0jk}\end{aligned}$$

$$\beta_{mjk} = \gamma_{m0k} \;\; \forall m \in \{1, \ldots, 4\}$$

Level 3:

$$\gamma_{00k} = \pi_{000} + \nu_{00k}$$

$$\gamma_{0nk} = \pi_{0n0} \ \ \forall n \in \{1, \ldots, 23\}$$

$$\gamma_{m0k} = \pi_{m00} \ \ \forall m \in \{1, \ldots, 4\}$$

In the next stage an All Levels Model was specified and fit which included all level 2 predictor variables and all level 3 predictor variables in order to determine their effects on Grade (as well as the variation between grades) when none of the slope coefficients on the level 1 predictor variables were allowed to vary between DLs or Lectures and none of the slope coefficients on the level 2 predictor variables were allowed to vary between Lectures. The slope coefficients on the DummyDLSize variables ($\pi_{0n0}$ for n $\in \{1, \ldots, 6\}$) as well as the slope coefficient on LecSize ($\pi_{001}$), though to a lesser degree given the speculation around active learning and class size above, are the primary focus of this study and will be used to answer the main research question.

## All Levels Model

Level 1:

$$\text{Grade}_{ijk} = \beta_{0jk} + \beta_{1jk} * \text{Male}_{ijk} + \beta_{2jk} * \text{LecStart}_{ijk} + \beta_{3jk} * \text{Units}_{ijk} + \beta_{4jk} * \text{GPA}_{ijk} + \varepsilon_{ij}$$

Level 2:

$$\begin{aligned}\beta_{0jk} = \gamma_{00k} + \sum_{n=1}^{6} \gamma_{0nk} * [\text{DummyDLSize}]_{jk} + \sum_{n=7}^{14} \gamma_{0nk} * \text{DL[Time]}_{njk} + \gamma_{015k} * \text{ROS}_{jk} + \gamma_{016k} \\ * \text{Winter}_{jk} + \gamma_{017k} * \text{Spring}_{jk} + \gamma_{018k} * \text{SevB}_{jk} + \gamma_{019k} * \text{SevC}_{jk} + \gamma_{020k} \\ * \text{Mean_Male}_{jk} + \gamma_{021k} * \text{Mean_LecStart}_{jk} + \gamma_{022k} * \text{Mean_Units}_{jk} + \gamma_{023k} \\ * \text{Mean_GPA}_{jk} + u_{0jk}\end{aligned}$$

$$\beta_{mjk} = \gamma_{m0k} \ \ \forall m \in \{1, \ldots, 4\}$$

Level 3:

$$\gamma_{00k} = \pi_{000} + \pi_{001} * \text{LecSize}_{k} + \nu_{00k}$$

$$\gamma_{0nk} = \pi_{0n0} \ \ \forall n \in \{1, \ldots, 23\}$$

$$\gamma_{m0k} = \pi_{m00} \ \ \forall m \in \{1, \ldots, 4\}$$

In the final stage of this process a series of models were specified and fit that sequentially randomized the slope coefficients of the level 1 predictor variables while also allowing them to co-vary with the level 2 constant error term ($u_{0jk}$). Based on a Chi-Squared test of the difference in Deviance between each of these models and the Individual Model with 26 additional degrees of freedom these models were a significantly better fit than the Individual Model [9, 47-49]. However, the model with the lowest Deviance and therefore the best fit while still being computable (models involving error terms in GPA slope coefficients did not run properly in the analysis) was the one which allowed the slope coefficient on Units to vary between DLs and co-vary with the level 2 constant error term. This is the Full Model.

## Full Model

Level 1:

$$\mathrm{Grade}_{ijk} = \beta_{0jk} + \beta_{1jk} * \mathrm{Male}_{ijk} + \beta_{2jk} * \mathrm{LecStart}_{ijk} + \beta_{3jk} * \mathrm{Units}_{ijk} + \beta_{4jk} * \mathrm{GPA}_{ijk} + \varepsilon_{ij}$$

Level 2:

$$\beta_{0jk} = \gamma_{00k} + \sum_{n=1}^{6} \gamma_{0nk} * [\mathrm{DummyDLSize}]_{jk} + \sum_{n=7}^{14} \gamma_{0nk} * \mathrm{DL}[\mathrm{Time}]_{njk} + \gamma_{015k} * \mathrm{ROS}_{jk} + \gamma_{016k} * \mathrm{Winter}_{jk} + \gamma_{017k} * \mathrm{Spring}_{jk} + \gamma_{018k} * \mathrm{SevB}_{jk} + \gamma_{019k} * \mathrm{SevC}_{jk} + \gamma_{020k} * \mathrm{Mean_Male}_{jk} + \gamma_{021k} * \mathrm{Mean_LecStart}_{jk} + \gamma_{022k} * \mathrm{Mean_Units}_{jk} + \gamma_{023k} * \mathrm{Mean_GPA}_{jk} + u_{0jk}$$

$$\beta_{mjk} = \gamma_{m0k} \ \ \forall m \in \{1, 2, 4\}$$

$$\beta_{3jk} = \gamma_{30k} + u_{3jk}$$

Level 3:

$$\gamma_{00k} = \pi_{000} + \pi_{001} * \mathrm{LecSize}_k + \nu_{00k}$$

$$\gamma_{0nk} = \pi_{0n0} \ \ \forall n \in \{1, \ldots, 23\}$$

$$\gamma_{m0k} = \pi_{m00} \ \ \forall m \in \{1, \ldots, 4\}$$

# Results

## Null Model

```
Log likelihood = -28631.756                    Prob > chi2       =          .

------------------------------------------------------------------------------
       Grade |      Coef.   Std. Err.      z    P>|z|     [95% Conf. Interval]
-------------+----------------------------------------------------------------
       _cons |   2.927973   .0329148    88.96   0.000     2.863461    2.992484
------------------------------------------------------------------------------

------------------------------------------------------------------------------
  Random-effects Parameters  |   Estimate   Std. Err.     [95% Conf. Interval]
-----------------------------+------------------------------------------------
Lecture: Identity            |
```

| | Estimate | Std. Err. | [95% Conf. | Interval] |
|---|---|---|---|---|
| var(_cons) | .0780536 | .0133515 | .05582 | .1091429 |
| DL: Identity | | | | |
| var(_cons) | .0036307 | .0017262 | .0014299 | .009219 |
| var(Residual) | .7944743 | .007743 | .7794425 | .8097961 |

Deviance = 57263.51

Table 2: Null Model Results

As seen in Table 2, the majority of variation (variance) in Grade is within DLs. However, with an $ICC_3$ of 0.0891, 8.91% of this variation exists between Lectures. Surprisingly, with an $ICC_2$ of 0.00414 only 0.414% of this variation exists between DLs within a given Lecture.

# Individual Model

Log likelihood = -20988.049 Prob > chi2 = 0.0000

| Grade | Coef. | Std. Err. | z | P>\|z\| | [95% Conf. | Interval] |
|---|---|---|---|---|---|---|
| GPA | 1.222879 | .0105282 | 116.15 | 0.000 | 1.202245 | 1.243514 |
| Units | -.0014798 | .0001696 | -8.73 | 0.000 | -.0018122 | -.0011475 |
| Male | .2237692 | .0099797 | 22.42 | 0.000 | .2042093 | .2433291 |
| LecStart | .0277263 | .0095328 | 2.91 | 0.004 | .0090423 | .0464103 |
| _cons | -.8303578 | .0491421 | -16.90 | 0.000 | -.9266746 | -.734041 |

| Random-effects Parameters | Estimate | Std. Err. | [95% Conf. | Interval] |
|---|---|---|---|---|
| Lecture: Identity | | | | |
| var(_cons) | .082535 | .0138644 | .0593816 | .1147162 |
| DL: Identity | | | | |
| var(_cons) | .0030051 | .0010969 | .0014695 | .0061454 |
| var(Residual) | .4547685 | .0046035 | .4458348 | .4638812 |

Deviance = 41976.1

Table 3: Individual Model Results

With a decrease in Deviance of 15287.41 from the Null Model and only four additional parameters (the fixed slope coefficients on the level 1 predictor variables) and thus, degrees of freedom in a Chi-Squared difference test of Deviance, the Individual Model is clearly (and unsurprisingly) a significantly better fit than the Null Model at the 95% level [9, p. 47-49]. The constant refers to the average value of Grade for a student with 0 on all of the predictor variables which includes Units and GPA so it has little physical meaning. In the future it might be helpful to grand mean center all or some of the continuous variables in these models (predictor and outcome). In terms of the slope coefficients, as seen in Table 3 all four of these are significant at the 95% level. As would be expected, a higher GPA is associated with a higher Grade such that, after controlling for all other predictor variables, a one point GPA increase leads to a 1.22 point increase in Grade on average. Also as expected, after controlling for all other predictor variables males get an average of 0.22 points higher on their Grades than females. Somewhat surprisingly, after controlling for all other predictor variables taking more Units prior to taking one of the courses studied leads to a lower Grade by an average of 0.0015 points per additional unit. Finally, being in the later lecture has a slightly positive effect on Grade of 0.028 points on average after controlling for all other predictor variables. The inclusion of these variables explained 38.7% of the variation within DLs but virtually none of the variation between DLs or Lectures.

## DL Model

```
Log likelihood = -20972.726                     Prob > chi2       =     0.0000

------------------------------------------------------------------------------
       Grade |      Coef.   Std. Err.      z    P>|z|     [95% Conf. Interval]
-------------+----------------------------------------------------------------
         GPA |   1.227061    .010744   114.21   0.000     1.206003    1.248118
       Units |  -.0014299   .0001732    -8.26   0.000    -.0017692   -.0010905
        Male |   .2236275   .0101829    21.96   0.000     .2036694    .2435856
    LecStart |   .0280326   .0096319     2.91   0.004     .0091546    .0469107
       RlySm |   .0753005   .2509686     0.30   0.764    -.4165889    .5671898
          Sm |  -.0493415   .0575736    -0.86   0.391    -.1621837    .0635006
         Lit |  -.0116786   .0340511    -0.34   0.732    -.0784174    .0550603
         Med |  -.0028365   .0164004    -0.17   0.863    -.0349807    .0293078
          Lg |  -.0288956   .0177588    -1.63   0.104    -.0637021     .005911
       RlyLg |  -.1261818   .1285011    -0.98   0.326    -.3780394    .1256758
         DL8 |   .0213201   .0180086     1.18   0.236     -.013976    .0566162
       DL105 |   .0088873   .0163125     0.54   0.586    -.0230846    .0408592
      DL1667 |   .0183755   .0162252     1.13   0.257    -.0134252    .0501762
      DL1917 |   .0578219   .0189071     3.06   0.002     .0207647     .094879
      DL1233 |    .010318    .023371     0.44   0.659    -.0354883    .0561243
      DL1234 |   .0724646   .1530654     0.47   0.636    -.2275382    .3724673
      DL1358 |  -.0451149   .1364339    -0.33   0.741    -.3125205    .2222906
      DL1542 |   .2257341   .2143656     1.05   0.292    -.1944147    .6458829
         ROS |   .0390543   .0488505     0.80   0.424    -.0566908    .1347995
      Winter |  -.1193747   .0796571    -1.50   0.134    -.2754997    .0367502
      Spring |  -.1396715   .0796439    -1.75   0.079    -.2957707    .0164277
        SevB |  -.1151129   .0803064    -1.43   0.152    -.2725106    .0422849
        SevC |   .0178901   .0824689     0.22   0.828    -.1437461    .1795262
    Mean_GPA |  -.0659132   .0556159    -1.19   0.236    -.1749184    .0430921
  Mean_Units |  -.0006831   .0008043    -0.85   0.396    -.0022595    .0008934
   Mean_Male |  -.0714489   .0581666    -1.23   0.219    -.1854534    .0425556
Mean_LecSt~t |   .0031765   .0697596     0.05   0.964    -.1335498    .1399028
       _cons |  -.4657486   .1950079    -2.39   0.017    -.8479571   -.0835401
------------------------------------------------------------------------------

------------------------------------------------------------------------------
  Random-effects Parameters  |   Estimate   Std. Err.     [95% Conf. Interval]
-----------------------------+------------------------------------------------
Lecture: Identity            |
                  var(_cons) |   .0740704   .0124821      .0532358    .1030591
-----------------------------+------------------------------------------------
DL: Identity                 |
                  var(_cons) |   .0023712   .0010639      .0009841    .0057134
-----------------------------+------------------------------------------------
               var(Residual) |   .4547544   .0046034      .4458209    .4638669
------------------------------------------------------------------------------
```

Deviance = 41945.45

Table 4: DL Model Results

With a decrease in Deviance of 30.65 and 23 additional parameters (the fixed slope coefficients on the level 2 predictor variables) the DL Model actually is not a significantly better fit than the Individual Model (the critical value for 95% confidence with 23 degrees of freedom is 35.2). On top of this, as seen in Table 4, while all level 1 predictor variable slope coefficients are significant and have similar qualitative interpretations as they did in the Individual Model the only level 2 predictor variable with a significant slope coefficient in this model is DL1917. According to this model (which should be rejected anyway) after controlling for all other predictor variables students in a DL that starts at 7:10pm get a grade that is an average of 0.058 points higher than those in a DL that starts at 2:10pm.

## All Levels Model

```
Log likelihood = -20972.074                     Prob > chi2       =     0.0000

------------------------------------------------------------------------------
       Grade |      Coef.   Std. Err.      z    P>|z|     [95% Conf. Interval]
-------------+----------------------------------------------------------------
         GPA |   1.227063    .010744   114.21   0.000     1.206005     1.24812
       Units |  -.0014288   .0001732    -8.25   0.000    -.0017682   -.0010895
        Male |   .2236302   .0101829    21.96   0.000     .2036721    .2435883
    LecStart |   .0280317   .0096319     2.91   0.004     .0091536    .0469098
       RlySm |    .072188   .2509753     0.29   0.774    -.4197144    .5640905
          Sm |  -.0501558   .0575757    -0.87   0.384    -.1630022    .0626906
         Lit |  -.0129173   .0340648    -0.38   0.705    -.0796832    .0538485
         Med |   -.003304   .0164044    -0.20   0.840    -.0354561    .0288481
          Lg |  -.0286788   .0177581    -1.61   0.106    -.0634839    .0061264
       RlyLg |  -.1255607   .1284958    -0.98   0.328    -.3774078    .1262865
         DL8 |   .0217779   .0180124     1.21   0.227    -.0135257    .0570815
       DL105 |   .0088699    .016312     0.54   0.587    -.0231011     .040841
      DL1667 |   .0183824   .0162247     1.13   0.257    -.0134174    .0501823
      DL1917 |   .0582191   .0189097     3.08   0.002     .0211569    .0952814
      DL1233 |   .0109098   .0233759     0.47   0.641    -.0349061    .0567256
      DL1234 |   .0734042   .1530619     0.48   0.632    -.2265917    .3734001
      DL1358 |  -.0431836   .1364405    -0.32   0.752    -.3106021     .224235
      DL1542 |   .2273682   .2143665     1.06   0.289    -.1927825    .6475188
         ROS |    .040198    .048859     0.82   0.411    -.0555639    .1359598
      Winter |  -.0936457   .0820851    -1.14   0.254    -.2545296    .0672382
      Spring |  -.1045504   .0846753    -1.23   0.217    -.2705109    .0614102
        SevB |  -.1601563   .0887771    -1.80   0.071    -.3341563    .0138437
        SevC |  -.0748388   .1150071    -0.65   0.515    -.3002485     .150571
    Mean_GPA |  -.0655204   .0556125    -1.18   0.239    -.1745189    .0434781
  Mean_Units |  -.0006747   .0008042    -0.84   0.401    -.0022509    .0009015
   Mean_Male |  -.0706097   .0581677    -1.21   0.225    -.1846163    .0433969
Mean_LecSt~t |   .0017713   .0697666     0.03   0.980    -.1349687    .1385114
     LecSize |  -.0010458   .0009115    -1.15   0.251    -.0028323    .0007408
       _cons |  -.1371507   .3457896    -0.40   0.692    -.8148859    .5405846
------------------------------------------------------------------------------

------------------------------------------------------------------------------
  Random-effects Parameters  |   Estimate   Std. Err.     [95% Conf. Interval]
-----------------------------+------------------------------------------------
Lecture: Identity            |
                  var(_cons) |   .0727532   .0122668      .0522797    .1012446
-----------------------------+------------------------------------------------
DL: Identity                 |
```

| | | | | |
|---|---|---|---|---|
| var(_cons) | .0023702 | .0010639 | .0009833 | .0057129 |
| var(Residual) | .4547553 | .0046034 | .4458218 | .4638679 |

Deviance = 41944.15

Table 5: All Levels Model Results

This model should also be rejected compared to the Individual model based on a Chi-Square difference test of their respective Deviances. Note that the coefficient on LecSize is also non-significant.

## Full Model

Log likelihood = -20957.181 Prob > chi2 = 0.0000

| Grade | Coef. | Std. Err. | z | P>\|z\| | [95% Conf. | Interval] |
|---|---|---|---|---|---|---|
| GPA | 1.226349 | .0107375 | 114.21 | 0.000 | 1.205304 | 1.247394 |
| Units | -.0015146 | .0001982 | -7.64 | 0.000 | -.001903 | -.0011262 |
| Male | .2235161 | .0101695 | 21.98 | 0.000 | .2035843 | .2434478 |
| LecStart | .0298841 | .0096264 | 3.10 | 0.002 | .0110166 | .0487515 |
| RlySm | .0820947 | .2515333 | 0.33 | 0.744 | -.4109015 | .5750908 |
| Sm | -.0515137 | .0578509 | -0.89 | 0.373 | -.1648994 | .0618721 |
| Lit | -.0142982 | .0342685 | -0.42 | 0.677 | -.0814633 | .0528669 |
| Med | -.0010056 | .0164929 | -0.06 | 0.951 | -.033331 | .0313198 |
| Lg | -.0264086 | .0179396 | -1.47 | 0.141 | -.0615696 | .0087523 |
| RlyLg | -.0646958 | .1374504 | -0.47 | 0.638 | -.3340936 | .2047021 |
| DL8 | .0199664 | .0182058 | 1.10 | 0.273 | -.0157162 | .055649 |
| DL105 | .0070199 | .0164019 | 0.43 | 0.669 | -.0251273 | .039167 |
| DL1667 | .0165593 | .0163457 | 1.01 | 0.311 | -.0154776 | .0485963 |
| DL1917 | .0568451 | .0191035 | 2.98 | 0.003 | .0194028 | .0942873 |
| DL1233 | .0083055 | .0236493 | 0.35 | 0.725 | -.0380463 | .0546573 |
| DL1234 | .073694 | .16263 | 0.45 | 0.650 | -.245055 | .392443 |
| DL1358 | -.0514849 | .1371075 | -0.38 | 0.707 | -.3202107 | .217241 |
| DL1542 | .2315272 | .2134239 | 1.08 | 0.278 | -.186776 | .6498304 |
| ROS | .0360103 | .0496379 | 0.73 | 0.468 | -.0612782 | .1332989 |
| Winter | -.0901174 | .0822769 | -1.10 | 0.273 | -.2513772 | .0711424 |
| Spring | -.1014241 | .0848522 | -1.20 | 0.232 | -.2677314 | .0648833 |
| SevB | -.161285 | .0889842 | -1.81 | 0.070 | -.3356908 | .0131208 |
| SevC | -.0752568 | .1152804 | -0.65 | 0.514 | -.3012021 | .1506886 |
| Mean_GPA | -.04957 | .0562378 | -0.88 | 0.378 | -.1597942 | .0606541 |
| Mean_Units | -.000632 | .0008108 | -0.78 | 0.436 | -.002221 | .0009571 |
| Mean_Male | -.0732751 | .0585716 | -1.25 | 0.211 | -.1880732 | .0415231 |
| Mean_LecSt~t | -.0070305 | .0707156 | -0.10 | 0.921 | -.1456306 | .1315695 |
| LecSize | -.00104 | .0009136 | -1.14 | 0.255 | -.0028307 | .0007506 |
| _cons | -.1776046 | .3474094 | -0.51 | 0.609 | -.8585145 | .5033052 |

| Random-effects Parameters | Estimate | Std. Err. | [95% Conf. | Interval] |
|---|---|---|---|---|
| Lecture: Identity | | | | |
| var(_cons) | .0730491 | .0123157 | .0524936 | .1016537 |
| DL: Unstructured | | | | |
| var(Units) | 5.75e-06 | 1.27e-06 | 3.73e-06 | 8.86e-06 |
| var(_cons) | .0390461 | .0043755 | .0313468 | .0486365 |

```
           cov(Units,_cons) |  -.0004612   .0000604     -.0005796   -.0003428
-----------------------------+------------------------------------------------
              var(Residual) |   .4498889   .0045861      .4409896    .4589679
------------------------------------------------------------------------------
```

Deviance = 41914.36

Table 6: Full Model Results

With a decrease in Deviance of 61.74 compared to the Individual Model and 26 additional parameters (23 fixed regression coefficients for the level 2 predictor variables, one from the fixed slope coefficient for LecSize, the variance in the error term for the regression coefficient on Units and the covariance between this error term, and the level 2 constant error term) the Full Model is a significantly better fit than the Individual Model. As seen in Table 6, all level 1 predictor variable slope coefficients are significant and have similar qualitative interpretations as they did in the Individual Model. The only other significant slope coefficient is the one on DL1917 which has the same qualitative interpretation as it did in the DL Model. Slightly more of the variation within DLs is explained by this model compared to the Individual Model and a bit of the variation between Lectures has been explained as well but the variation between DLs within a given Lecture has actually increased (even relative to the Null Model) by quite a bit.

# Discussion and Conclusions

The slope coefficients on no class size (DL or Lecture) variables were significant at the 95% level indicating that perhaps DL (and Lecture) size do not impact student understanding in these settings (at least to a point since a class of say, 100, would be a different story that is beyond the scope of this study). Therefore, in light of this study there is less of a reason than there otherwise might have been to focus attention or resources on reducing the size of introductory university physics classes. However, it might be possible to design and implement a Force Concept Inventory (FCI) style assessment for these courses to better gauge student understanding instead of using course grades as a proxy for understanding which has some merit but is still nowhere near a perfect indicator of student understanding when it comes to underlying concepts. Doing a controlled experiment which accounts for the possible effect of teaching assistants (TAs) on student understanding would be quite helpful as well, especially if done using an FCI type of evaluation, since oftentimes TAs teach DLs in unique ways. Really though the biggest confounding factor here is the fact that a very small portion of the variance in Grades even exists at the DL level to begin with which means level 2 predictor variables will not have much of an impact on Grades regardless of anything else. Perhaps this means that DLs are so well organized that the differences between them do not have much of an impact on student learning but another possibility is that the primary determinants in student Grades, quizzes and exams created and administered by course instructors, may not correspond that well to what is being taught in DLs. This potential problem of the validity of these Grades is something that should be strongly considered and examined further.

# Limitations and Future Work

One obvious limitation besides the skewness and ceiling and floor effects described previously is the missing observations. Beyond these missing data points systematically excluding graduate students, first quarter freshmen, first quarter transfer students, and students who take these courses over the summer as discussed earlier it is assumed that the rest are random and do not systematically exclude any other groups of students who take these courses. This is a claim which could be empirically tested. Another major limitation is the fact that many of these data points belong to the same student. If this is because they took all three courses (or two of them) this effect should mostly be accounted for through the Winter and Spring dummy variables (which are merely controls and not something under investigation), though the interactions between these dummy variables and other predictor variables in the study was not examined and in the future it may be useful to do a separate analysis for each of the three courses under study. However, if there are students who have repeated a course (which there almost certainly are) causing there to be multiple observations for the same student for a single course the additional unaccounted for correlation between these observations could have an effect on the results. One way to address this in the future is to include a dummy variable for students who have repeated a course (or even an ordinal variable for the number of times a student has repeated the course).

Finally, there are several additional factors which should be controlled for and could be in future research, provided access to the proper data sets is granted. For instance, using high school GPA or students' grades in the University's introductory calculus or chemistry courses either along with or instead of GPA could help. Not controlling for the TAs who run the DLs is also an issue that could be addressed in the future, though the question there then becomes how to do so. Additionally, in the future students' race/ethnicity should be controlled for as well.